\documentclass{elsart}

\usepackage{epsfig}
\usepackage{amsmath}
\usepackage{amssymb}

\begin{document}

\begin{frontmatter}

\title{System-size dependence
of strangeness production in high-energy A+A collisions and
percolation of strings }

\author{C.~H\"{o}hne\corauthref{cor1}}
\ead{c.hoehne@gsi.de} \corauth[cor1]{Corresponding author.}
\address{Gesellschaft f\"{u}r Schwerionenforschung (GSI), Darmstadt, Germany}
\author{F.~P\"{u}hlhofer}
\address{Fachbereich Physik der Philipps-Universit\"{a}t Marburg, Germany}
\author{R.~Stock}
\address{Fachbereich Physik der Johann Wolfgang Goethe-Universit\"{a}t Frankfurt, Germany}


\begin{abstract}

We argue that the shape of the system-size dependence of
strangeness production in nucleus-nucleus collisions can be
understood in a picture that is based on the formation of clusters
of overlapping strings. A string percolation model combined with a
statistical description of the hadronization yields a quantitative
agreement with the data at $\sqrt{s_{NN}}=17.3$~GeV. The model is
also applied to RHIC energies.

\end{abstract}


\end{frontmatter}


\section{Introduction}

The enhanced production of strangeness-carrying hadrons in
nucleus-nucleus in comparison to p+p collisions is a well-known
phenomenon. It has been observed over a wide range of collision
energies from threshold to $\sqrt{s_{NN}}=200$~GeV. The
explanation for it depends on the energy regime. Here, we
concentrate on the range of the CERN SPS ($\sqrt{s_{NN}}\sim
7-20$~GeV), in particular on the top energy. As opposed to the
purely hadronic scenario discussed for the low energies
\cite{dunlop,wang}, there is experimental evidence for a change to
a partonic phase in the early stage of nucleus-nucleus collisions,
e.g. from the energy dependence of the
$\langle\text{K}\rangle$/$\langle\pi\rangle$ and
$\langle\Lambda\rangle/\langle\pi\rangle$ ratios
\cite{na49_energy,na49_lambda} or of the mean transverse momenta
of the emitted particles \cite{christoph_sqm04}. Due to the high
energy density the interacting matter apparently traverses the
deconfinement transition \cite{marek}. Whether the
strange-particle content is determined in the partonic phase
\cite{marek}, or during hadronization, as suggested in
\cite{stock,becattini,becattini2,pbm-wetterich}, or to what extent
it is modified during the hadronic expansion phase, is still an
open question.

An attempt of a quantitative description of all particle yields is
made by statistical models \cite{becattini,statmodel}. The
reaction system is approximated by a multi-particle state of
maximum probability. Its properties are obtained from the rules of
statistical theory. It is equivalent to a system in
thermodynamical equilibrium characterized in the grand-canonical
case by the parameters $T$, $\mu_{B}$ and $V$ (temperature,
baryonic potential, and reaction volume). In this picture,
strangeness enhancement in large systems is ascribed to a volume
dependence of relative strangeness production. In small systems,
like p+p, strangeness production is suppressed as a consequence of
the requirement of strict
flavor conservation (canonical strangeness suppression)
\cite{rafelski,redlich}. Aside from details, this effect is
independent of whether a hadronic or partonic scenario is
considered.

An alternative to the macroscopic picture inspired by
thermodynamics is given by microscopic reaction theories, e.g. the
transport models UrQMD \cite{urqmd} or VENUS \cite{venus}. They
follow the individual nucleon trajectories and simulate a
nucleus-nucleus reaction by a sequence of color-string formation
and decay processes. At high
collision densities the strings start overlapping. This could
give rise to an increased strength of the color field and, as a
consequence, to enhanced strange-particle production.

The CERN-NA49 Collaboration has recently published the yields of a
series of strangeness-carrying particles (charged kaons,
$\Lambda$, $\phi$, in addition charged pions for normalization) in
central C+C and Si+Si collisions at the top SPS energy
($\sqrt{s_{NN}}=17.3$~GeV) \cite{na49size}. Together with earlier
data for p+p, S+S and Pb+Pb the systematics of strangeness
production as a function of the size of the collision system or,
equivalently, of the number $N_{\rm part}$ of participating
nucleons was obtained. For all individual particles, including the
$\phi$-meson, the pion-normalized yields show a steep increase
starting at p+p followed by a saturation at about 60 participating
nucleons.

Comparing this behavior to the results of statistical theories
based on the effect of canonical strangeness suppression  a
serious discrepancy is observed. The theory
\cite{rafelski,redlich} predicts a much faster rise with a
saturation at a value roughly six times below the observed one.
This conclusion is based on the usual assumption that the reaction
volume is proportional to the number of the colliding nucleons.
The discrepancy can not be removed by a different choice of the
proportionality factor nor by the assumption that only a fixed
fraction of nucleons contributes.

In this Letter, the relation between the reaction volume $V$ and
the number of colliding nucleons is investigated more carefully.
Starting from a microscopic model of a nucleus-nucleus collision
the density of the individual strings is calculated. It is then
assumed that due to the statistical overlap of the corresponding
strings clusters of highly excited and strongly interacting matter
are formed. The technique applied is a percolation model. In
general, these clusters consist of subgroups of the participating
nucleons. Only with heavy collision partners a single cluster
comprising nearly all participants is formed. The clusters are
then assumed to hadronize independently. The resulting particle
compositions are calculated from the statistical model, i.e. the
effect of canonical strangeness suppression is used, though at a
different level. The fact that for small collision systems several
hadronizing volumes are formed instead of one large cluster leads
to a softening of the system size dependence of relative
strangeness production, and the theoretical results come into
agreement with the experimental data.

The model is not necessarily limited to applying a statistical
description of the hadronization process. As mentioned above, the
overlap of strings may by itself lead to an enhanced strangeness
production. This alternative, as it is e.g.\,implemented in the
RQMD model \cite{sorge} is not considered here.

\section{The model}
\label{sec_model}

The main purpose of the model is to calculate the system-size
dependence of the relative strangeness production in nucleus-nucleus
collisions. As indicated in the Introduction, the collision
process is separated into two independent steps:
\begin{itemize}
\item formation of coherent clusters by coalescence
         of strings (percolation part);
\item hadronization of the clusters, simulated by a statistical
         description, i.e.\,in essence the canonical suppression function
         (hadronization part).
\end{itemize}
Any effects of interactions in the final hadronic expansion stage
are neglected.

The microscopic model VENUS \cite{venus} is used to obtain the
density of the individual NN collisions during the penetration of
the two nuclei. The calculations are performed for the systems
studied in \cite{na49size}: p+p (minimum-bias trigger) and
approximately central C+C, Si+Si, S+S and Pb+Pb collisions at a
center-of-mass energy $\sqrt{s_{NN}}=17.3$~GeV. In this paper, the
mean number $N_{\rm wound}$ of wounded nucleons calculated from a
Glauber model is used to characterize the system size. For the
central collision systems the values were taken from
\cite{na49size}.

\begin{figure}[]
\epsfig{file=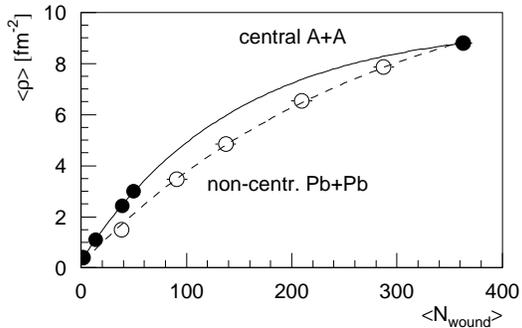,width=7cm} \vspace{-0.2cm}
\caption{\label{rho_vs_Nwound} Mean area density
$\langle\rho\rangle$ of NN collisions in the transverse plane
calculated with the VENUS model in dependence on the number of
wounded nucleons $N_{\rm wound}$ for near-central A+A collisions
of different mass number A (solid line). The dashed line holds for
non-central Pb+Pb collisions with varying centrality discussed in
section~4.}
\end{figure}

The longitudinal dimension of the reaction zone in the
center-of-mass system is given by the Lorentz-contracted diameter
of one of the colliding nuclei. For the lighter systems with
$N_{\rm wound}<60$, which are the ones of interest for the shape
of the system size dependence of strangeness production, this
diameter is below 1~fm. It is not reasonable to subdivide this
longitudinal dimension any further, but to integrate over it. With
this simplification, one obtains a 2-dimensional density
distribution as well as a mean area density $\langle\rho\rangle$
of NN interactions in the transverse plane. The latter quantity is
shown in figure \ref{rho_vs_Nwound}. A smooth function fitted to
the calculated systems is used for interpolation. The shape of the
density distribution, averaged over many collisions, turns out to
be rather independent of the system under study, as long as
near-central collisions are considered. In the following, a common
density profile is used.

\begin{figure}[]
\epsfig{file=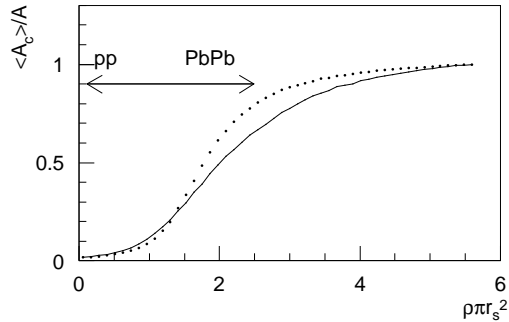,width=7cm} \vspace{-0.2cm}
\caption{\label{A_vs_rho} The normalized mean cluster size
$\langle A_{c}\rangle/A$ (including "clusters" comprising only
single strings) as a function of the the string area density
$\rho\cdot\pi r_{s}^{2}$. Dotted line: uniform density
distribution; solid line: realistic profile (from VENUS). The
range of mean densities from minimum bias p+p to central Pb+Pb
collisions is indicated by the arrow for $r_{s}=0.3$~fm.}
\end{figure}

The next step is a percolation calculation. The strings are given
a transverse area $A_{s}=\pi r_{s}^{2}$, with an effective string
radius $r_{s}$. They are distributed over the transverse area $A$
of the reaction zone (obtained as the geometrical overlap of
the two nuclei represented by spheres with the 10\%-density radius).
The assumption is that as soon as the strings start to
overlap they form clusters. As shown in figure \ref{A_vs_rho}, the
mean cluster size $\langle A_{c}\rangle$ (or equivalently the
normalized value $\langle A_{c}\rangle/A$) rises steeply at a
critical string density and reaches a saturation state at
densities which are typical for central collisions of heavy nuclei.
In this limiting case, only one single cluster containing all
participants is formed. The behavior described is called a
percolation phase transition. Results are
shown for uniformly distributed strings (dotted line) and for
strings distributed according to a density profile as obtained
from VENUS  (solid line), the latter leading to a softer transition.
The simulations of figure \ref{A_vs_rho} are actually made for
an overlap area $A=16$~fm$^{2}$ representative for semi-central
C+C collisions, but aside from details like finite-size effects
\cite{satz} they have general validity.

\begin{figure}[]
\epsfig{file=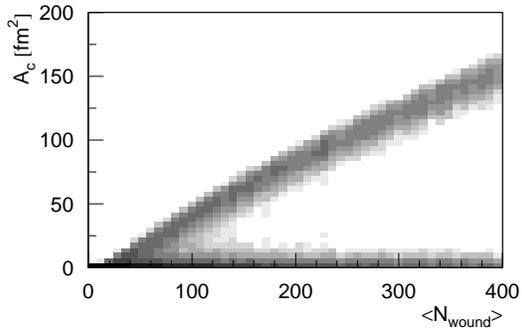,width=7cm} \vspace{-0.2cm}
\caption{\label{A_vs_Nwound} Distribution of cluster areas
including single strings as a function of $N_{\rm wound}$
($r_{s}=0.3$~fm, simulation with realistic density profile). The
grey-shade coding is on a logarithmic scale.}
\end{figure}

Combining the results from figure \ref{rho_vs_Nwound} and figure
\ref{A_vs_rho} the mean cluster size $\langle A_{c}\rangle$ can be
obtained for all collision systems, i.e.\,as a function of $N_{\rm
wound}$. For the final results, however, it is important to use
not only the mean value, but to take into account the
variation of the cluster sizes within a single collision. The
probability density distribution of $A_{c}$ for various systems is
presented in figure \ref{A_vs_Nwound}.

The next task is to treat the hadronization of the clusters.
Here, the only goal is to describe the relative strangeness
production. Assuming that a cluster of overlapping strings can be
considered as a coherent entity, the statistical model is applied
to obtain the strangeness suppression factor $\eta$ (as defined in
\cite{rafelski,redlich}) and its dependence on the volume $V$ of
the hadronizing system. One has two choices: either to calculate
the strangeness content in the partonic phase, or in the final
hadronic phase. The former implies that it is not changed during
hadronization, the latter uses the hadron gas model as an
effective hadronization model. It has been shown \cite{rafelski}
that both assumptions yield very similar results. The following
calculations are based on the model for a partonic phase as
described by Rafelski and Danos \cite{rafelski}.

The first step is to transform the cluster sizes in the transverse
plane to 3-dimensional hadronization volumes $V_{h}$. An obvious
way to extrapolate $A_{c}$ to a volume $V_{h}$ is a simple scaling
relative to the values for a N+N collision characterized by
$V_{0}$ and $A_{0}=\pi r_{s}^{2}$:
\begin{equation}
V_{h}=A_{c} \cdot \frac{V_{0}}{\pi r_{s}^{2}} \label{formula_Vn}
\end{equation}
The hadronization volume for p+p, $V_{0}$, is usually taken to be
of the order of the nucleon volume. Here we leave it as adjustable
parameter fixed by the experimental strangeness enhancement
between p+p and Pb+Pb. The quantity $\frac{V_{0}}{\pi r_{s}^{2}}$
could be interpreted as composed of a string length and a factor
that accounts for the expansion until hadronization.

\begin{figure}[]
\epsfig{file=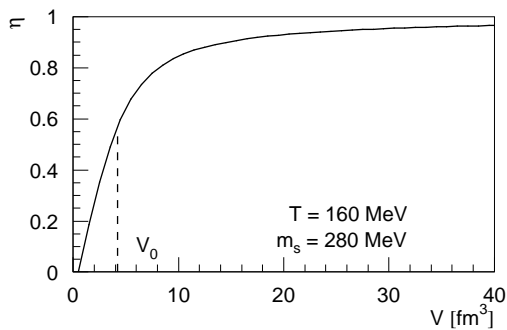,width=7cm}
\vspace{-0.2cm} \caption{\label{s-suppression} Strangeness
suppression factor $\eta$ in dependence on the reaction volume
calculated according to \cite{rafelski} in a quark phase with
$T=160$ MeV and a strange-quark mass $m_{s}=280$ MeV.
$V_{0}$=4.2~fm$^{3}$ is indicated by the dashed line. }
\end{figure}

The relative strangeness abundance of a quark phase in a given
cluster volume $V_{h}$ can then be calculated following
\cite{rafelski}. The result is shown in figure \ref{s-suppression}
for a parameter choice suggested in \cite{rafelski}. The final
task is to combine all the steps described and to compare the
results to the experimental data.

\section{Comparison with experiment}

Experimentally, the total relative strangeness production as
e.g.\,described by the Wroblewski factor $\lambda_{s}=\frac{2
\langle s\overline{s} \rangle }{\langle
u\overline{u}+d\overline{d}\rangle }$ is not directly accessible.
Therefore, it will be approximated by the measured fraction of the
main strangeness carriers to pions:
\begin{equation}
E_{s}=\frac{\langle \Lambda \rangle + 2(\langle
K^{+}\rangle+\langle K^{-}\rangle)}{\langle\pi\rangle}
\end{equation}
with
$\langle\pi\rangle=\frac{3}{2}(\langle\pi^{+}\rangle+\langle\pi^{-}\rangle)$.
The model of the previous section is used here only for predicting
the variation of $E_{s}$  with system size. It is assumed that
$E_{s}$ is proportional to the strangeness suppression factor
$\eta$ averaged over all clusters in a collision:
\begin{equation}
\label{formula_es} E_{s}(N_{\rm wound})=a \cdot
\langle\eta(V_{h})\rangle
\end{equation}
Strictly speaking, $\langle\eta(V_{h})\rangle$ is a
volume-weighted mean, averaged over many collisions for a specific
system.

\begin{figure}[]
\epsfig{file=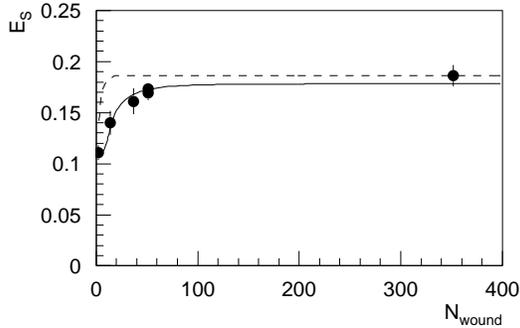,width=7cm}
\vspace{-0.2cm} \caption{\label{es} System-size dependence of
strangeness production. The data for p+p and for near-central A+A
collisions at $\sqrt{s_{NN}}=17.3$~GeV from NA49 \cite{na49size}
are given by dots. The solid line is from the model described in
the text. The dashed line is the result of the statistical model
with the common assumption $V=\frac{V_{0}}{2}N_{\rm wound}$ with
$V_{0}= 4.2$ fm$^{3}$.}
\end{figure}

Figure \ref{es} shows a comparison between the NA49 data for the
relative strangeness abundance as a function of system size and
the results of the model. The parameters are: $a=0.18$, $T$ and
$m_{s}$ as given in figure \ref{s-suppression}, $r_{s}=0.3$ fm and
$V_{0}=4.2$~fm$^{3}$. A perfect agreement is obtained.

The difference to the dashed line in figure \ref{es} demonstrates
the improvement over the assumption of a proportionality between
the reaction volume and $N_{\rm wound}$, that has usually been
made in the literature so far. If, as in this Letter, the full
cluster size distribution is taken into account, the rise of the
system size dependence is softened as required by the data.

In total, only a few free parameters are needed for the model. For
the effective string radius $r_{s}$ values in the range of 0.2-0.3
fm are motivated by perturbative QCD calculations \cite{satz}. A
value of $r_{s}=0.3$~fm is chosen here, similarly as has been used
by other percolation calculations
\cite{satz,percolation,percolation-pt-fluc,percolation-mult}. A
variation of $r_{s}$ changes the steepness of the strangeness
increase for small systems. It becomes weaker the smaller $r_{s}$
is. The parameter $V_{0}$ effects the overall strangeness
enhancement from p+p to central Pb+Pb. The scaling parameter $a$
(eq. \ref{formula_es}) is well determined by the data. Decreasing
the strange quark mass to $m_{s}=150$~MeV would require a
decreased $V_{0}$ of 3~fm$^{3}$ in order to fit the data while
leaving $r_{s}$=0.3~fm. Using uniformly distributed strings
instead of the density distribution from VENUS or slightly
different definitions of the overlap area $A$ leads only to
negligible effects.

\section{Further applications of the model}

In the preceding section the focus lay on the total strangeness
production in (near-)central collisions between nuclei of equal,
but varying mass at $\sqrt{s_{NN}}=17.3$~GeV. It suggests itself
to apply the model to other collision geometries and energies.

\begin{figure}[]
\epsfig{file=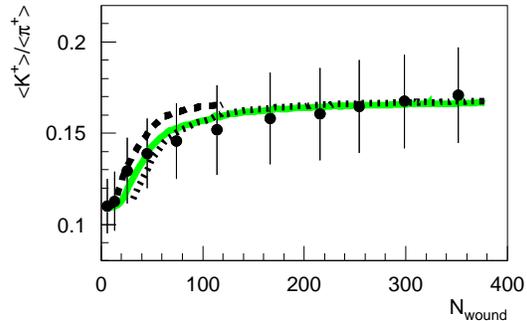,width=7cm}
\vspace{-0.2cm} \caption{\label{es_rhic} Strangeness production in
centrality-controlled collisions between heavy nuclei. Data
points: midrapidity $\langle K^{+}\rangle/\langle\pi^{+}\rangle$
yields for Au+Au at $\sqrt{s_{NN}}=200$~GeV from PHENIX
\cite{phenix}. Solid line: calculated $E_{s}$ for this system,
rescaled to fit the data. Short dashed line: $E_{s}$ for Pb+Pb at
$\sqrt{s_{NN}}=17.3$~GeV demonstrating the expected similarity
between RHIC and SPS energies. Long dashed line: prediction for
Cu+Cu at  $\sqrt{s_{NN}}=200$~GeV.}
\end{figure}

Assuming that the same strangeness production mechanism holds for
top SPS and RHIC energies, the model can be applied to the higher
energies. In order to adjust the simulation to the slightly
reduced strangeness enhancement at $\sqrt{s_{NN}}=200$~GeV
compared to top SPS energies, the volume parameter was increased
to $V_{0}=4.6$~fm$^{3}$. All other parameters except for the
scaling parameter $a$ were taken from the calculations before; the
dependence of $\langle\rho\rangle$ on $N_{\rm wound}$ for Au+Au
was extracted from VENUS. Assuming that the $\langle
K^{+}\rangle/\langle\pi^{+}\rangle$ ratio represents the total
relative strangeness production, figure \ref{es_rhic} shows the
model calculation compared to the ratio of midrapidity yields from
PHENIX \cite{phenix}. Again, the system-size dependence of
relative strangeness production is described well.

As only the elementary hadronization volume was slightly changed
to optimize the fit to the RHIC data, a qualitatively similar
system-size dependence for centrality dependent Pb+Pb collisions
at top SPS energies is expected. A corresponding calculation is
shown by the short dashed line in figure \ref{es_rhic}. The slight
difference at low $N_{\rm wound}$ mainly results from the
different volumina $V_{0}$ used for SPS and RHIC. Compared to the
calculation for near-central A+A shown in figure \ref{es}, the
increase is slightly weakened; this is caused by the different
relation between the collision density and the number of wounded
nucleons (see figure \ref{rho_vs_Nwound}). The relative
strangeness production at a certain density $\rho$ is further
reduced compared to near-central A+A, since the larger transverse
areas $A$ in peripheral Pb+Pb slightly shift the percolation
transition to higher densities \cite{satz}. As Cu+Cu data at
$\sqrt{s_{NN}}=200$~GeV will be available in the near future, a
calculation for this system is added by the long dashed line. If
any, only a slightly higher relative strangeness production at the
same $N_{\rm wound}$ is expected.

The model presented here is in principle also applicable to
further interesting observables. A similar percolation ansatz has
been used successfully for the explanation of other system-size
dependencies \cite{percolation}, e.g.\, of the $\langle
p_{t}\rangle$ fluctuations as observed at SPS and RHIC
\cite{percolation-pt-fluc}, or to the dependence of the mean
multiplicity per event on $N_{\rm wound}$ \cite{percolation-mult}.
In particular the explanation of the fluctuations could be related
to the fact that in small systems a large number of differently
sized clusters exists, while in p+p only a single string and in
Pb+Pb only one large cluster is present. From this observation
also increased fluctuations in the relative strangeness production
would be expected for small systems such as central C+C, Si+Si, or
peripheral Pb+Pb collisions.

\section{Summary}

In this Letter, the system-size dependence of relative strangeness
production in nucleus-nucleus collisions is described by a model
that is based on a statistical hadronization picture including, as
an essential part, the effect of canonical strangeness
suppression, the latter describing the dependence on the
hadronization volume. This volume is not, as usual, assumed to be
proportional to the number of nucleons participating in the
collision, but calculated from a microscopic reaction model for
each individual collision type from the coalescence of strings.
According to this model, several clusters exist in smaller
collision systems, while in central Pb+Pb basically one large
cluster is created comprising all participating nucleons. This
effect based on the different geometry of the systems is shown to
be essential for reproducing the data. The model is applicable in
the energy regime $\sqrt{s_{\rm NN}}\geq 17$~GeV.

The system size dependent growth of high energy density regions
(the clusters formed by overlapping strings in the model presented
here) within the primordial interaction volume can serve as a
mechanism to understand the transition from a canonical to a
grand-canonical description of hadronization, which also occurs
with the growth of the statistical ensemble volume. If we assume,
not unplausibly, that the single clusters decay
quantum-mechanically coherent, quantum number conservation occurs
globally over the whole volume as is characteristic for quantum
system decays. The outcome thus resembles the effect of the
chemical potential employed in the grand-canonical model, which
enforces quantum number conservation only on average, over the
entire hadronizing volume. This effect causes the observed
strangeness enhancement in larger collision systems.

{\bf Acknowledgements:} We thank our NA49 colleagues, in
particular ~V.~Friese, ~M.~Gazdzicki and ~P.~Seyboth, for valuable
discussions.




\end{document}